\documentclass[9pt,twocolumn,twoside]{pnas-new}
\templatetype{pnasresearcharticle} % Choose template
\usepackage{hyperref}

\title{Is the ''Dark Comet" 2005 VL$_1$ the Venera 2 Spacecraft?}

\author[1,2]{Abraham Loeb}
\author[1]{Richard Cloete}

\affil[1]{Institute for Theory \& Computation, Harvard University\\60 Garden Street\\Cambridge MA 02138, USA}

\leadauthor{Loeb}

\significancestatement{
This paper shows that 2005 VL$_1$, originally classified as a ''dark comet", may in fact be the Venera 2 spacecraft launched in 1965, illustrating how some near-Earth objects (NEOs) may be remnants of past space missions. Identifying such space debris among NEOs has key implications for planetary defense, mission planning, and the interpretation of small Solar System bodies. This finding underscores the challenge of distinguishing real comets or asteroids from unrecognized technological fragments, highlighting the need for continued observational and orbital analyses. By demonstrating that technological space debris contributes to newly discovered objects, this work broadens our understanding of both NEOs and human-made debris in the inner Solar System, with important implications for the upcoming Legacy Survey of Space and Time (LSST) of the Vera C. Rubin Observatory.
}

\authorcontributions{A.L defined the goals of the study, identified Venera 2 as the possible counterpart of 2005 VL$_1$, calculated the Solar radiation pressure acceleration of Venera 2 and wrote the paper. R.C. performed the orbital calculations and derived the numerical results that substantiated this association.}
\authordeclaration{The authors declare no commercial interests in this study.}
\correspondingauthor{\textsuperscript{2}To whom correspondence should be addressed. E-mail: aloeb\@cfa.harvard.edu}


\begin{abstract}
Recently, Ref. \citep{Seligman24} identified a population of near-Earth objects (NEOs) that exhibit statistically-significant non-gravitational accelerations with no coma, and labeled them ''dark comets". Here, we show that one of these objects, 2005 VL$_1$, was at closest approach to Earth in November 1965 when the Venera 2 spacecraft was launched to explore Venus. The observed $H$ magnitude of 2005 VL$_1$ is consistent with a high reflectance from the  full surface of Venera 2 including its Solar panels. As known for Venera 2, 2005 VL$_1$ arrived within a short distance 
from Venus in February 1966, a highly improbable coincidence ($\lesssim 1\%$) for the orbital phase of a near-Earth object that does not target a close approach to Venus. Indeed, 2005 VL$_1$'s orbital parameters are very similar to the reported values for Venera 2. Given the area-to-mass ratio of Venera 2, we show that 2005 VL$_1$'s non-gravitational acceleration and negligible transverse acceleration match the values expected from Solar radiation pressure. 
\end{abstract}

\dates{This manuscript was compiled on \today}
\doi{\url{www.pnas.org/cgi/doi/10.1073/pnas.XXXXXXXXXX}}

\begin{document}

\maketitle
\thispagestyle{firststyle}
\ifthenelse{\boolean{shortarticle}}{\ifthenelse{\boolean{singlecolumn}}{\abscontentformatted}{\abscontent}}{}

% \firstpage{52}

\section{Introduction}
Following the Luna 1 spacecraft to the Moon in 1959, the Solar System was polluted with upper stages of human-made rockets, which appear today as near-Earth objects (NEOs). For example, 2020 SO \citep{Battle24} was identified as the Centaur upper stage used to launch the 1966 Surveyor 2 spacecraft, which was found to be pushed by Solar radiation pressure upon its discovery by the Pan-STARRS observatory on September 17, 2020, nearly three years after the same observatory discovered 1I/`Oumuamua~\citep{Williams} - which also featured anomalous non-gravitational acceleration~\citep{Micheli}. Other examples of human-made technological objects include XL8D89E\footnote{\url{https://www.projectpluto.com/pluto/mpecs/xl8d89e.htm}}, 2018 AV2\footnote{\url{https://www.projectpluto.com/pluto/mpecs/2018av2.htm}}, and 2023 NM\footnote{\url{https://www.projectpluto.com/pluto/mpecs/2023nm.htm}}. In addition, J002E3\footnote{\url{https://cneos.jpl.nasa.gov/news/news134.html}} is thought to be the third-stage Saturn S-IVB booster from Apollo 12, having been identifed to have an almost identical orbit.

Recently, Ref. \citep{Seligman24} listed data on 14 near-Earth objects (NEOs) that exhibit statistically significant non-gravitational accelerations. The objects were labeled as ''dark comets'' because there is no sign of coma around any of them.

Here, we suggest that one of these objects, 2005 VL$_1$, might be the Soviet Venera 2 spacecraft, launched to explore Venus on November 12, 1965~\footnote{\url{https://nssdc.gsfc.nasa.gov/nmc/spacecraft/display.action?id=1965-091A}}. Radio communication with Venera 2 was lost at closest approach to Venus in February 1966, and its whereabouts were unknown afterwards.

\section{Orbital Calculation}

% https://ssd.jpl.nasa.gov/tools/sbdb_lookup.html#/?sstr=2005%20VL1

We use the orbital parameters of 2005 VL$_1$ as listed in Table 1 of Ref. \citep{Seligman24} (originally reported in Ref. \citep{Seligman23}) and integrate the orbit back in time. 

Remarkably, 2005 VL$_1$ was at closest approach to Earth around the time of the Venera 2 mission’s launch to explore Venus on November 12, 1965. The inferred perihelion of 2005 VL$_1$ is 0.69 au, similar to the 0.72 au orbital radius of Venus. As expected for Venera 2, our calculation of the orbit of 2005 VL$_1$ implies that it arrived within 
%$2\times 10^{-4}~{\rm au}$ 
a short distance from Venus in February 1966, a highly improbable coincidence for the orbital phase of a natural comet that does not target a close encounter with Venus. Since the orbital period of Venus is 225 days, the probability for an NEO arriving near Venus at the right phase within a few days is $\lesssim 1\%$.  

We use the orbit of 2005 VL$_1$ with eccentricity 0.225, perihelion distance of 0.691 au and inclination of 0.235$^\circ$ which were measured with high precision~\footnote{\url{https://ssd.jpl.nasa.gov\/tools\/sbdb\_lookup.html\#\/?sstr=2005\%20VL1}}, 
and agree with the eccentricity and semimajor axis of Venera 2. We ignore the discrepancy with the reported inclination of Venera 2 ($i=4.29^{\circ}$) which was likely a measurement error owing to the poor radio communication with the spacecraft or a result of orbital maneuvering or a gravitational kick from Venus~\footnote{\url{https://en.wikipedia.org/wiki/Venera\_2}}.

\section{Solar Radiation Pressure as the Source of the Non-Gravitational Acceleration of 2005 VL$_1$}

The Venera 2 spacecraft can be approximated as a cylinder plus rectangular Solar panels with a total mass of $M=0.963\times 10^6~{\rm g}$, a length of $\sim 3.3$~m~and a width of $\sim4.4$~m\footnote{\url{https://nssdc.gsfc.nasa.gov/nmc/spacecraft/display.action?id=1965-091A}}. Adopting a fiducial cross-sectional area of $S\approx 10^5~{\rm cm^2}$ that reflects a fraction $R$ of the Solar radiation impinging on it at a distance $r$ from the Sun, the resulting non-gravitational acceleration away from the Sun is,
\begin{equation}
A = - \frac{2 R S L_{\odot}}{4\pi\,r^2 M c}
  = -\,4.9\times 10^{-9}\,\frac{R}{\bigl({r}/{\mathrm{1~au}}\bigr)^{2}}\,~~\frac{\mathrm{au}}{\mathrm{d}^2}\,.
  \label{accel}
\end{equation}
This non-gravitational push away from the  Sun declines inversely with heliocentric distance squared, $\propto r^{-2}$, like the Sun's gravity. It amounts to a reduction in the Sun's gravitational acceleration by a fraction of $1.86\times 10^{-5}R$. We use this simple reduction in the Sun's gravity when calculating the shifts relative to the expected gravitational trajectory of 2005 VL$_1$.

The observed $H$ magnitude of 26.45 implies a high reflectance, possibly from the Solar panels or the metallic enclosure of Venera 2.\footnote{{\url{https://cneos.jpl.nasa.gov/tools/ast\_size_est.html}}}

The only statistically significant component of the reported non-gravitational acceleration of 2005 VL$_1$ is out of the ecliptic plane. Given 2005 VL$_1$'s orbital inclination of $i=0.25^{\circ}=4.4\times 10^{-3}$~radians, the predicted out-of-plane component of the Solar radiation acceleration for Venera 2 is,
\begin{equation}
    A_3 = A \sin i 
    = -2.1 \times 10^{-11}\,\frac{R} 
      {\bigl({r/{1~{\rm au}}}\bigr)^{2}}~~
      \,\frac{\mathrm{au}}{\mathrm{d}^2}\,.
\end{equation}

For high reflectance, this expected result for Venera 2 is comparable to the inferred value of $A_3= - (2.41\pm 0.395) \times 10^{-11}[r/{\rm 1~au}]^{-2} ~{\rm au~d^{-2}}$ for 2005 VL$_1$ in Table 2 of Ref. \citep{Seligman24}. The transverse acceleration $A_2$ is consistent with zero, as expected for Solar radiation pressure. The radial acceleration in the ecliptic plane should be $A_1\approx A$ in equation~(\ref{accel}), which is consistent with the value $A_1= - (8.30\pm 7.59) \times 10^{-10}[r/{\rm 1~au}]^{-2} ~{\rm au~d^{-2}}$ reported in Ref.~\citep{Seligman23}. Since the Venera 2 spacecraft is not spherically symmetric, the solar radiation force perpendicular to any reflecting surface could also have a non-radial component. This additional component is difficult to model, as it requires knowledge of the orientation of the spacecraft and its rotation rate. 

To demonstrate consistency of the expected radiation pressure on Venera 2 with the non-gravitational shifts of 2005 VL$_1$, we compared the expected astrometric shifts from radiation pressure directly to the reported data in Ref.~\citep{Seligman23}.

\section{RA and DEC Shifts by Solar Radiation Pressure}

We integrated the orbit of 2005 VL$_1$ based on the reported orbital parameters and calculated the residual Right Ascension (RA) and Declination (DEC) for an orbit influenced by radiation pressure on Venera 2 with the Sun's gravity reduced by a fraction
$1.86\times 10^{-5}R$ compared to an orbit with the full Sun's gravity.

\begin{figure*}[ht]
    \centering
    \includegraphics[width=0.8\textwidth]{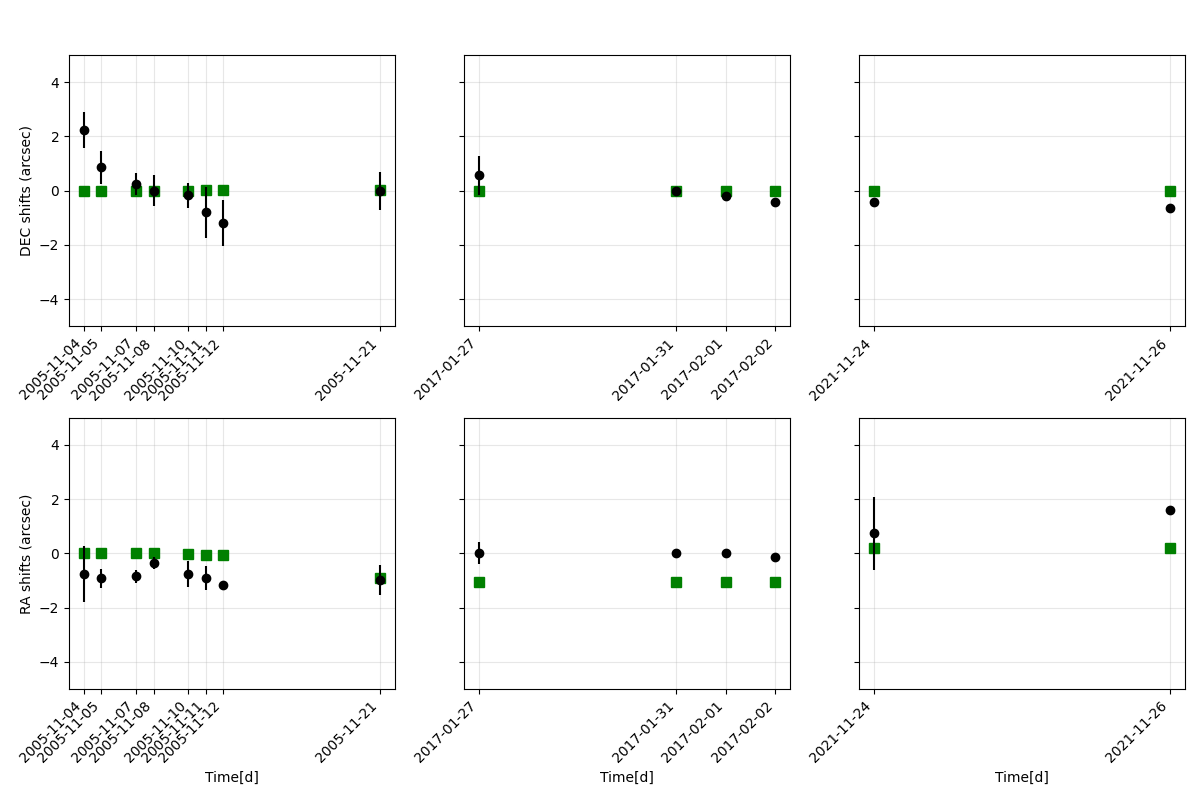}
    \caption{Shift in astrometric fits (green points) that are required to bring the Declination (DEC) and Right Ascension (RA) of Venera 2's orbit with $A= - 1.5 \times 10^{-9}[r/{\rm 1~au}]^{-2} ~{\rm au~d^{-2}}$ (or $R\sim 0.3$), to a Keplerian orbit with gravity only. The calculated shifts are compared to the reported shifts (black points with error bars) based on the daily average of the data collected during the dates when measurements were made for 2005 VL$_1$~\citep{Seligman23}. The solar radiation force could also have a transverse component for a non-spherical object, which may improve the fit to the data. }
    \label{fig:img}
\end{figure*}

Figure~\ref{fig:img} shows the shifts in DEC and RA as a result of Solar radiation pressure acting on Venera 2, compared to the data reported for 2005 VL$_1$ at the bottom of Figure 2 of Ref.~\citep{Seligman23}. The original data was reported in three boxes, each extending over periods of time between November 4 to 19, 2005, January 27 to February 2, 2017 and January 24 to 26, 2022. We start the comparison at the first data point on November 4, 2005, where by the definition the shifts should be set to zero. Subsequently, we compare the daily average of the shift measured for 2005 VL$_1$~\citep{Seligman23} to our expectations for Venera 2 with $A= - (1.5) \times 10^{-9}[r/{\rm 1~au}]^{-2} ~{\rm au~d^{-2}}$, corresponding to $R\sim 0.3$.

The approximate agreement between the observed and expected data for the DEC and RA  shifts implies that 2005 VL$_1$'s non-gravitational acceleration is consistent with the expected Solar radiation pressure on Venera 2, considering the possible unmodeled non-radial force components owing to the non-sphericity of Venera 2.

\section{Discussion}
We showed that 2005 VL$_1$, cataloged as a ''dark comet'' by Refs. \citep{Seligman23} and \citep{Seligman24} based on its non-gravitational accelerations and lack of a coma, might be the Venera 2 spacecraft. Indeed, 2005 VL$_1$ was at closest approach to Earth around the time when Venera 2 spacecraft was launched in November 12, 1965, and 2005 VL$_1$  arrived within 
%$2\times 10^{-4}~{\rm au}$ 
a short distance from Venus on February 27 1966 just as Venera 2. Such orbital alignment with Earth and Venus is highly improbable ($\lesssim 1\%$) for a comet, as a natural object does not typically follow trajectories matching a spacecraft’s deliberate path between Earth and another planet.

The $H$ magnitude of 26.45 for 2005 VL$_1$ is consistent with the expected size of Venera 2 for a high albedo. The out-of-plane non-gravitational acceleration and negligible transverse acceleration of 2005 VL$_1$ are consistent with Solar radiation pressure given its area-to-mass ratio, as demonstrated in Figure~\ref{fig:img}. Since the Venera 2 spacecraft is not spherically symmetric, additional components of the solar radiation force - which are difficult to model because of its complicated shape and unknown rotation rate, could yield better agreement with the data. Additional data would help to validate our argument and measure any unmodeled force components.

On January 2, 2025, the Minor Planet Center(MPC) announced~\footnote{\url{https://minorplanetcenter.net/mpec/K25/K25A38.html}} a new NEO asteroid labeled 2018 CN41 with an orbital period of 1.53 years. Within less than 17 hours, an editorial notice was issued by the MPC~\footnote{\url{https://minorplanetcenter.net/mpec/K25/K25A49.html}} on the deletion of 2018 CN41 from its database since the object was the {\it Tesla Roadster} car, launched on February 6, 2018, as a dummy payload on the {\it Falcon Heavy} rocket's first flight. This car is known to be orbiting the Sun on the same eccentric orbit reported for 2018 CN41.

It is conceivable that other "dark comets" are technological space debris similar to 2005 VL$_1$, in which case their inferred non-gravitational acceleration is not the result of invisible cometary evaporation. For example, 2016 GW$_{221}$ share nearly identical inclination, perihelion and aphelion and closest distance from Earth on April 2, 1964, as the Zond 1 spacecraft to Venus~\footnote{\url{https://en.wikipedia.org/wiki/Zond\_1}}. To accommodate such objects, the label "dark comets" should be replaced by the more appropriate label "objects with anomalous non-gravitational acceleration (OANGAs)," because their nature may be different from that of the familiar class of comets. Some OANGAs could be either the products of human technologies over the past 66-year history of the modern space age or products of extraterrestrial civilizations over the past billions of years. Interstellar objects in the latter category could have been either trapped by gravitational interactions with massive planets like Jupiter~\citep{Napier} or designed to settle in the inner Solar System.

\acknow{This work was supported in part by the Galileo Project at Harvard University.}

\showacknow % Display the acknowledgements section

% Bibliography
\bibliography{refs.bib}

\begin{thebibliography}{1}

\bibitem{Seligman24}
Seligman DZ, et~al. (2024) Two distinct populations of dark comets delineated by orbits and sizes.
\newblock {\em Proceedings of the National Academy of Sciences} 121(51):e2406424121.

\bibitem{Battle24}
{Battle} A, et~al. (2024) {Challenges in Identifying Artificial Objects in the Near-Earth Object Population: Spectral Characterization of 2020 SO}.
\newblock {\em PSJ} 5(4):96.

\bibitem{Williams}
{Williams} GV, et~al. (2017) {Minor Planets 2017 SN\_33 and 2017 U1}.
\newblock {\em Central Bureau Electronic Telegrams} 4450:1.

\bibitem{Micheli}
{Micheli} M, et~al. (2018) {Non-gravitational acceleration in the trajectory of 1I/2017 U1 ('Oumuamua)}.
\newblock {\em Nature} 559:223--226.

\bibitem{Seligman23}
{Seligman} DZ, et~al. (2023) {Dark Comets? Unexpectedly Large Nongravitational Accelerations on a Sample of Small Asteroids}.
\newblock {\em PSJ} 4(2):35.

\bibitem{Napier}
{Napier} KJ, {Adams} FC, {Batygin} K (2021) {On the Fate of Interstellar Objects Captured by Our Solar System}.
\newblock {\em PSJ} 2(6):217.

\end{thebibliography}

\end{document}